\documentstyle[12pt]{article}

\newcommand{\ct}[1]{\cite{#1}}
\newcommand{\bi}[1]{\bibitem{#1}}
\newcommand{\lleq}[1]{\label{eq:#1}}
\newcommand{\beql}[1]{\begin{equation} \lleq{#1}}
\newcommand{\req}[1]{(\ref{eq:#1})}
\newcommand{\bm}[1]{\mbox{\boldmath $#1$}}
\newcommand{\sav}[1]{\left\{ #1 \right\}}

\def\beq{\begin{equation}}     \def\eeq{\end{equation}}
\def\beqs{\[}                  \def\eeqs{\]}
\def\beqa{\begin{eqnarray}}    \def\eeqa{\end{eqnarray}}
\def\ov{\over}
\def\lg{\left\langle}    \def\rg{\right\rangle}
\def\lgs{\langle}        \def\rgs{\rangle}
   
\def\non{\nonumber}

\def\Nev{N_{\rm ev}}

\def\rmf{{f}}

\newcommand{\regpreprint}[4]
{\noindent\begin{minipage}[t]{\textwidth}\begin{center}
\framebox[\textwidth]
{$\rule[6mm]{0mm}{0mm}$ \raisebox{1.3mm}
{#1{Institut f\"ur Theoretische Physik der Universit\"at Regensburg}}
}

\vspace{2mm}    \rule{\textwidth}{0.2mm}\\
\vspace{-4mm}   \rule{\textwidth}{1pt}
\mbox{ }    #2    \hfill    #3   \mbox{ }\\
\vspace{-2mm}   \rule{\textwidth}{1pt}\\
\vspace{-4.2mm} \rule{\textwidth}{0.2mm}
\end{center}
\end{minipage}

\vspace{2mm}

\mbox{}\hfill hep-ph/9309256

\vspace{10mm}
\begin{center} {#4}  \end{center}
}   

\input epsf

\begin{document}

\regpreprint{\large\bf}{September 1993}{TPR--93--25}
{\large\bf
Star integrals and unbiased estimators$^*$}

\vspace{5mm}
\begin{center}
H.\ C.\ Eggers\footnote{
Present address: Dept.\ of Physics, McGill University, Montr\'eal,
Qu\'ebec, Canada H3A 2T8\\
\mbox{}\qquad eggers@prism.physics.mcgill.ca}\\
{\it
Institut f\"ur Theoretische Physik, Universit\"at Regensburg,\\
D--93040 Regensburg, Germany  } \\
\mbox{ }\\
and \\
\mbox{ }\\
P.\ Lipa\footnote{
Present address: Inst.\ f\"ur Hochenergiephysik,
Nikolsdorfergasse 18, A--1050 Vienna, Austria\\
\mbox{}\qquad v2033dag@awiuni11.edvz.univie.ac.at} \\
{\it Department of Physics, University of Arizona, Tucson AZ 85721, USA  } \\
\mbox{}\\
\mbox{}\\
\end{center}

\begin{quote}
\centerline{ABSTRACT}
\small
We review in brief the development and implementation of
the Star integral, a tool yielding measurements of correlations
much superior to conventional methods. A version for
use in pion interferometry is explained. We also show how effects
of non-poissonian overall multiplicity distributions may be
eliminated if desired and quote results eliminating
statistical biases arising in correlation measurements within small samples.
\end{quote}

\vfill\noindent
$^*$Talk presented at the Cracow Workshop on Multiparticle Production,
April 1993.


\section{Point distributions}

Distributions are fundamental to almost any branch of the exact
sciences. A {\it point distribution} is characterized by the
fact that the object under scrutiny has no intrinsic structure or
size; it is a point rather than a field. The coordinates of this point
may be discrete or continuous, real or complex, embedded in a
single- or multidimensional space. Physically, whenever the objects
(``particles'') have a size small in comparison with the embedding
space, the assumption of a point distribution is justified.

Typical examples of point distributions are galaxies in the
sky and pions in phase space; in these cases, the
embedding space is continuous. If $\bm{X}_i$ are the coordinates of
$N$ particles measured in a particular ``event''
(as measured by the detector, a region of the sky, a throw of $N$ dice,
\ldots), the number of such particles at the point $\bm{x}$ is
\beql{aas}
\hat\rho_1(\bm{x}) = \sum_{i_1=1}^N \delta(\bm{x} - \bm{X}_{i_1}) \,.
\eeq
In general, the simultaneous behavior ( = correlation) of $q$ of
these particles for a given event $a$ is given by
\beql{ciaa}
\hat\rho_q^a(\bm{x}_1,\ldots,\bm{x}_q) =
\sum_{\scriptstyle i_1\neq i_2\neq\ldots\neq i_q \atop \scriptstyle =1}^{N}
\delta (\bm{x}_1-\bm{X}^a_{i_1})\,
\delta (\bm{x}_2-\bm{X}^a_{i_2})    \,\cdots\,
\delta (\bm{x}_q-\bm{X}^a_{i_q}) \,.
\eeq
Note that $\hat\rho_q$ is a nonnegative integer while the
coordinates $\bm{X}$ are continuous. Meaningful results are
extracted by averaging over many events, to yield the
$q$-tuple density
\beql{aau}
\rho_q = \lg \hat\rho_q^a \rg
       = \Nev^{-1} \sum_{a=1}^{\Nev} \hat\rho_q^a \,,
\eeq
which is pseudocontinuous (actually, a rational number
with denominator $\Nev$). Allowing the variables to range
over some restricted domain $\Omega$, one measures
{\it moments} of point distributions
\beql{aat}
\xi_q(\Omega) = \int_\Omega \rho_q(\bm{x}_1,\ldots,\bm{x}_q)
\, d\bm{x}_1 \ldots d\bm{x}_q
= \lg n^{[q]} \rg_\Omega
\eeq
which contain information on the correlation strength
in this particular region $\Omega$.
Here $n^{[q]} \equiv n!/(n-q)! = n(n{-}1)\cdots(n{-}q{+}1)$,
so that $\xi_q$ is actually a {\it factorial moment} of
the distribution.

Below, we explore one particular type of domain $\Omega$ which
leads to the so-called {\it Star integral}. The reason for
this particular choice of $\Omega$
is that it maximizes the amount of information extracted from
a given sample of events while restricting the amount of
computer time to a minimum \ct{Egg93a}.

\newpage

\section{Star integral moments}

Conventional measurements of correlations proceed first to discretize
the continuous variable $\bm{X}$ (``binning the data'') and then
to find $\xi_q$ by averaging over all events the counts
$n_m^{[q]}$ in every bin \ct{Bia86a},
\beql{aba}
\xi_q^{\rm conv} = \lg \sum_{{\rm bins}\ m} n_m^{[q]} \rg .
\eeq
By contrast,
the Star integral belongs to the class of {\it correlation integrals},
where distances between pairs of points
$X_{i_1 i_2} \equiv | \bm{X}_{i_1} - \bm{X}_{i_2} |$
are computed directly, before binning \ct{Lip92a}.
For the Star integral, the domain $\Omega$ is given by the sum
of all spheres of radius $\epsilon$ centered around each of the
$N$ particles in the event. The number of particles (``sphere count'')
within each of these spheres is, not counting the particle at
the center $\bm{X}_{i_1}$,
\beql{ciad}
\hat n(\bm{X}_{i_1},\epsilon)
\equiv
\sum_{i_2=1}^N \Theta(\epsilon - X_{i_1 i_2})
\,,\ \ \ \ \ \ i_2 \neq i_1 \,,
\eeq
and the factorial moment of order $q$ is
\beql{ciac}
\xi_q^{\rm Star}(\epsilon) = \lg
\sum_{i_1}
\hat n(\bm{X}_{i_1},\epsilon)^{[q-1]} \rg  .
\eeq
The obvious similarity to the conventional moment of Eq.\ \req{aba}
should not disguise the fact that $\sum_{i_1}$ is a sum over
particles rather than bins.

The Star factorial moment of Eq.\ \req{ciac} can be derived rigorously
\ct{Egg93a} from Eq.\ \req{ciaa} using for $\Omega$ the equivalent
implicit definition
\beql{abb}
\xi_q^{\rm Star}(\epsilon) =
\int \rho_q(\bm{x}_1,\ldots,\bm{x}_q) \,
\Theta_{12}\Theta_{13}\ldots\Theta_{1q} \,
d\bm{x}_1 \ldots d\bm{x}_q
\eeq
with the theta functions
$\Theta_{1j} \equiv \Theta(\epsilon - |\bm{x}_1 - \bm{x}_j|)$
restricting all $q{-}1$ coordinates $\bm{x}_j$ to within a distance
$\epsilon$ of $\bm{x}_1$.

Correlation measurements can be made either as a function of the sphere
size $\epsilon$ --- useful in looking for self-similarity and
fractal structure --- as a function of the distance from a fixed center
coordinate, as has been the case traditionally. This is implemented
in the ``differentials'' of Section 3 below.

The superiority of the Star integral moments of Eq.\
\req{abb} over the conventional $\xi_q^{\rm Conv}$ arises because
the artificial discretization inherent in the latter has two
bad effects \ct{Lip92a}. First, it leads to
instabilities in the measured quantities $n^{[q]}$ as the bin
size is varied and, second, it often sorts particles into different
bins when in fact they are quite close together, thus unwittingly
throwing away information. By contrast, $\xi_q^{\rm Star}$
is much more stable and has smaller errors, especially if the
coordinates in use live in a two- or three-dimensional space.

In the literature on galaxy correlations \ct{Pal84a}
and in the characterization of strange attractors \ct{GHP83},
an approach similar to our Star integral has been used for some time,
utilizing, however, not the factorial but the ordinary form
\beql{abd}
\lg {1\ov N^2} \sum_{i_1} \hat n(\bm{X}_{i_1},\epsilon)^{q-1} \rg .
\eeq
We believe that this is an {\it ad hoc} approximation to
the exact factorial form of Eq.\ \req{ciac}, necessarily breaking
down when $\hat n$ becomes small. Current measurements in these fields
may therefore be suffering from distortion at small $\epsilon$.

In order to eliminate, among other things, the overall multiplicity,
it has become customary in high energy physics to measure
{\it normalized factorial moments} \ct{Bia86a}. The denominator
used for such normalization should be made up of the uncorrelated
background, $\rho_1^q$. While it can be implemented in a number of
ways, we prefer the ``vertical'' normalization, in which $\rho_1^q$
is integrated over exactly the same domain $\Omega$ as the inclusive
density $\rho_q$ in the numerator. Thus for the Star integral, the
normalized moment is
\beql{cigc}
F_q^{\rm Star}(\epsilon) \equiv
{\xi_q^{\rm Star} \ov \xi_q^{\rm norm}  }  =
{
\int \rho_q(\bm{x}_1,\ldots,\bm{x}_q) \,
\Theta_{12}\Theta_{13}\ldots\Theta_{1q} \,
d\bm{x}_1 \ldots d\bm{x}_q
\ov
\int \rho_1(\bm{x}_1)\ldots\rho_1(\bm{x}_q) \,
\Theta_{12}\Theta_{13}\ldots\Theta_{1q} \,
d\bm{x}_1 \ldots d\bm{x}_q
} \,.
\eeq
We have shown rigorously \ct{Egg93a} that
the denominator $\xi_q^{\rm norm}$ is given by the following double
event average: with  $X_{i_1 i_2}^{ab} \equiv
| \bm{X}_{i_1}^{a} - \bm{X}_{i_2}^{b} |$ measuring the distance
between two particles {\it taken from different events}
$a$ and $b$,
\beql{evd}
\xi_q^{\rm norm}(\epsilon) =
\lg \sum_{i_1}
\lg \sum_{i_2}
\Theta(\epsilon - X_{i_1 i_2}^{ab})
\rg^{q-1}
\rg
\equiv
\lg
\sum_{i_1} \lg \hat n_b(\bm{X}_{i_1}^a,\epsilon)\rg^{q-1}
\rg ,
\eeq
where the outer event average and sum over $i_1$ are taken over the center
particle taken from event $a$, each of which is used as the center
of sphere counts $\hat n_b(\bm{X}_{i_1}^a,\epsilon)$ taken over other
events $b$ in the inner event average. We thus see the natural
emergence of the heuristic procedure of normalization known as
``event mixing'' \ct{Egg93a,Lip92a}.

Apart from the double event average and the appearance of the
ordinary power $q{-}1$ rather than the factorial power $[q{-}1]$,
the similarities between the numerator Eq.\ \req{ciac} and denominator
Eq.\ \req{evd} are obvious. Both do sphere counts around a given
center particle at $\bm{X}_{i_1}$; the numerator $\xi_q^{\rm Star}$
does so within the same event $a$, while the denominator
$\xi_q^{\rm norm}$ inserts this center particle into all other
events $b$ to perform a similar count there. This is shown
schematically in Figure 1.

\section{Cumulants and differentials}
\label{sec:cudif}

Cumulants are combinations of correlation functions constructed
in such a way as to become zero whenever any one or more
of the points $\bm{x}$ becomes statistically independent of the others.
This is done so as to strip away the combinatorial background from the
correlation measurements,
\beqa
\lleq{cua}
C_2(\bm{x}_1,\bm{x}_2) &=& \rho_2(\bm{x}_1,\bm{x}_2) -
\rho_1(\bm{x}_1)\rho_1(\bm{x}_2) \,, \\
\lleq{cub}
C_3(\bm{x}_1,\bm{x}_2,\bm{x}_3) &=& \rho_3(\bm{x}_1,\bm{x}_2,\bm{x}_3)
               -\ \rho_1(\bm{x}_1)\rho_2(\bm{x}_2,\bm{x}_3)
               -\ \rho_1(\bm{x}_2)\rho_2(\bm{x}_3,\bm{x}_1)\non\\
    & &\mbox{} -\ \rho_1(\bm{x}_3)\rho_2(\bm{x}_1,\bm{x}_2)
            +\ 2\,\rho_1(\bm{x}_1)\rho_1(\bm{x}_2)\rho_1(\bm{x}_3)
                 \quad \mbox{etc.}
\eeqa

\vspace*{12mm}
\begin{center}
\begin{minipage}[h]{60mm}
\epsfysize=60mm\epsfbox{eggersfig1.eps}
\end{minipage}
\end{center}

\begin{quote}
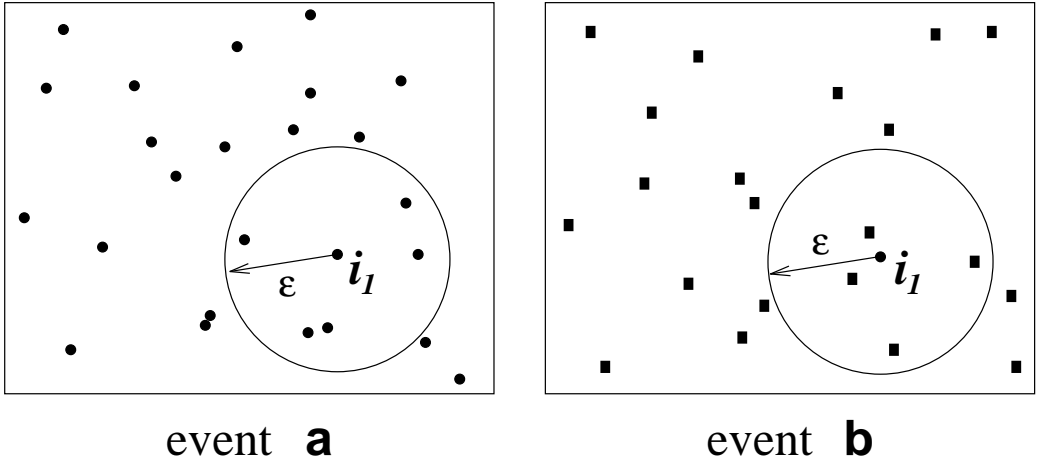
{\small {\bf Figure 1:} Sphere counts. On the left is shown a
typical event $a$, with the particles denoted as circles.
For each particle $i_1$ of $a$, count all other particles
within a sphere of radius $\epsilon$; this yields
$\hat n(\bm{X}_{i_1},\epsilon)$ of Eq.\ \req{ciad} used in the
numerator of the Star integral moments and cumulants.
On the left is shown a different event $b$, with particles
denoted as squares. For normalization and cumulants, the same
center particle  is inserted at $\bm{X}_{i_1}$ into event $b$
and a count performed to yield $\hat n_b(\bm{X}_{i_1},\epsilon)$.
Performing an event average over all $b$-events, one obtains the
normalization $\xi_q^{\rm norm}$ as in Eq.\ \req{evd} and cumulants
as in Eq.\ \req{cuo}.
}
\end{quote}

\vspace*{5mm}

\noindent
Integrating these over the Star integral domain, we can find the
normalized cumulants
\beql{cuh}
K_q^{\rm Star}(\epsilon) \equiv
{ \rmf_q(\epsilon) \over \xi_q^{\rm norm}(\epsilon) } \,,
\eeq
with
\beql{cui}
\rmf_q(\epsilon) \equiv
\int C_q(\bm{x}_1,\ldots,\bm{x}_q) \,
\Theta_{12}\Theta_{13}\ldots \Theta_{1q} \,
d\bm{x}_1\ldots d\bm{x}_q
\eeq
the unnormalized (Star) factorial cumulant. The latter can be written
entirely in terms of the sphere counts introduced previously;
writing in shorthand
\beqa
\lleq{cuja}
a &=& \sum_{j}\Theta(\epsilon - X_{ij}^{aa})
   = \hat n(\bm{X}_i^a,\epsilon), \ \ \ \  j\neq i \\
\lleq{cuj}
b &=& \sum_j \Theta(\epsilon - X_{ij}^{ab})
   = \hat n_b(\bm{X}_i^a,\epsilon) \,
\eeqa
and defining for convenience the ``$i$-particle cumulant''
$\hat\rmf_q(i)$ by
\beql{cuo}
\lg \sum_i \hat\rmf_q(i) \rg = \rmf_q \,,
\eeq
we find
\beqa
\lleq{cupa}
\hat\rmf_2(i) &=& a - \lgs b \rgs \,, \\
\lleq{cupb}
\hat\rmf_3(i)
&=& a^{[2]} - \lgs b^{[2]} \rgs - 2 a \lgs b \rgs
            + 2 \lgs b \rgs^2 \,,
\eeqa
etc. In Section \ref{sec:ubia} below, we shall show that these
$\hat\rmf_q$ as well as the normalization $\xi_q^{\rm norm}$ must be
corrected for a remaining statistical bias. This correction should
become important for small data samples.

Besides counting the number of combinations of $q$ particles
lying inside a sphere of radius $\epsilon$, a second useful
form for Star moments and cumulants are the so-called
{\it differential} moments: Here, one defines not only a maximum
distance $\epsilon_t$ but a minimum also, $\epsilon_{t-1}$
($t$ can define a sequence of such distances).
For a given combination of $q{-}1$ particles around a center
particle at $\bm{X}_{i_1}$, at least one of these must lie inside
the spherical shell bounded by radii $\epsilon_{t-1}$
and $\epsilon_t$, while the others are restricted only by the
maximum distance $\epsilon_t$. This is illustrated in Figure 2.

\vspace*{8mm}

\begin{center}
\begin{minipage}[h]{58mm}
\epsfysize=58mm\epsfbox{eggersfig2.eps}
\end{minipage}
\end{center}

\vspace{1mm}

\begin{quote}
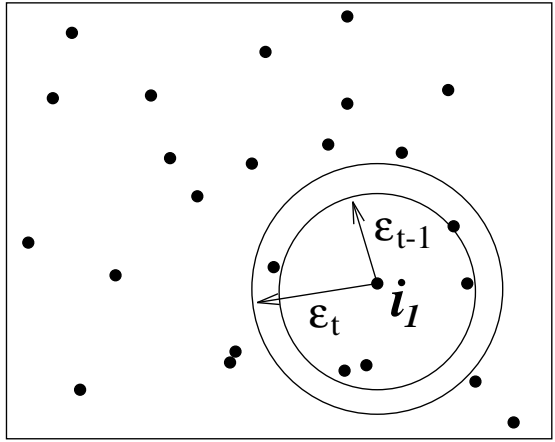
{\small {\bf Figure 2:} Sphere counts for differentials.
Given the center particle at $\bm{X}_{i_1}$, at least
one other particle must be in the shell bounded
by radii $\epsilon_{t-1}$ and $\epsilon_t$ to count.
For $q=2$, this reduces to subtracting the
sphere count for $\epsilon_{t-1}$ from that for
$\epsilon_t$. Higher orders are also easily calculated.
}
\end{quote}

\vspace{4mm}

This definition leads rigorously \ct{Egg93a} to simple and efficient
prescriptions for measurements. For $q=2$, the unnormalized
differential moment is, with
$\Delta\xi_q(\epsilon_t) = \lg \sum_{i_1}
     \Delta\hat\xi_q(i_1,\epsilon_t) \rg$
\beql{dde}
\Delta\hat\xi_2(i_1,t) =
\hat n(\bm{X}_{i_1}^a,\epsilon_t) -
\hat n(\bm{X}_{i_1}^a,\epsilon_{t-1})
\equiv a_t - a_{t-1} \,,
\eeq
the latter defining the shortened notation. For higher orders, we find
\beql{ddf}
\Delta\hat\xi_q(i_1,t) =
\hat n(\bm{X}_{i_1}^a,\epsilon_t)^{[q-1]} -
\hat n(\bm{X}_{i_1}^a,\epsilon_{t-1})^{[q-1]}
\equiv a_t^{[q-1]} - a_{t-1}^{[q-1]} \,,
\eeq
i.e.\ just the difference of $[q{-}1]$th factorial powers of two
sphere counts. Equivalent forms for the (differential) normalizations
$\Delta\hat\xi_q^{\rm norm}$ and differential cumulants $\Delta\hat\rmf_q$
are easily found, leading to the normalized differential
moments and cumulants
\beqa
\lleq{ddj}
\Delta F_q(t)
&=& {
\lg \sum_i  a_t^{[q-1]} - a_{t-1}^{[q-1]} \rg
\ov
\lg \sum_i
\lg b_t     \rg^{q-1}  -
\lg b_{t-1} \rg^{q-1}  \rg
} \,, \\
\lleq{ddk}
\Delta K_q(t)
&=& {
\lg \sum_i  \hat\rmf_q(i,\epsilon_t) -
                      \hat\rmf_q(i,\epsilon_{t{-}1}) \rg
\ov
\lg \sum_i
\lg b_t     \rg^{q-1}  -
\lg b_{t-1} \rg^{q-1}  \rg
} \,,
\eeqa
in obvious notation. The point is that these quantities can all
be measured in terms of the two types of sphere counts, $\hat n$
within the same $a$-event and $\hat n_b$ within the other $b$-events;
see Eqs.\ \req{cuja}--\req{cuj}.

\section{Eliminating effects of the overall multiplicity distribution}

If there are $N$ particles within the total phase space
(sky region) $\Omega_{tot}$ considered, the normalized factorial moment
for this whole region is
$F_q(\Omega_{tot}) =  \lgs N^{[q]} \rgs / \lg N \rg^q$,
which is unity only when the overall multiplicity distribution is
poissonian. All measurements of $F_q$, $K_q$ and their differentials
thus implicitly contain correlations arising from the non-poissonian
nature of the overall multiplicity distribution. This is as it
should be, of course, but it may sometimes be desirable to
eliminate this dependence on global effects (for example when the
multiplicity distribution is artificial,
such as in centrality cuts in heavy ion collisions).
One way of achieving this is to modify all the formulae of the preceding
sections by changing the event-by-event counts to
\beql{elb}
\hat\rho_1(\bm{x}) \, \longrightarrow \,
\hat h_1(\bm{x})
= {1\ov \hat N}  \sum_{i_1=1}^N \delta(\bm{x} - \bm{X}_{i_1})
= {                      \hat\rho_1(\bm{x})  \ov
     \int_{\Omega_{tot}} \hat\rho_1(\bm{x}) d\bm{x}  } \,,
\eeq
where $\hat N$ is the event's
multiplicity within the total domain $\Omega_{\rm tot}$, and
$\hat\rho_q$ to
\beqa
\lleq{elc}
\hat h_q(\bm{x}_1,\ldots,\bm{x}_q)
&=&
{1\ov \hat N^{[q]} }
  \sum_{\scriptstyle i_1\neq i_2\neq\ldots\neq i_q \atop \scriptstyle =1}^{N}
\delta (\bm{x}_1-\bm{X}_{i_1})\,
\delta (\bm{x}_2-\bm{X}_{i_2})    \,\cdots\,
\delta (\bm{x}_q-\bm{X}_{i_q}) \non\\
&=& {                     \hat\rho_q(\bm{x}_1,\ldots,\bm{x}_q)  \ov
      \int_{\Omega_{tot}} \hat\rho_q(\bm{x}_1,\ldots,\bm{x}_q)
        \, d\bm{x}_1 \ldots d\bm{x}_q  } \,.
\eeqa
The event average $h_q = \lgs \hat h_q \rgs$ satisfies the
requirements of a joint probability (normalization to unity and correct
projection properties).
These changes then propagate to yield, for example
\beql{eld}
F_q(\epsilon) =  \lg {1 \ov \hat N_a} \sum_{i_1}
                    { a^{[q-1]} \ov (\hat N_a-1)^{[q-1]}  } \rg
                 \left/
                 \lg {1 \ov \hat N_a} \sum_{i_1}
                      \lg b \ov \hat N_b \rg^{q-1} \rg
                 \right. \,,
\eeq
which yields $F_q(\Omega_{tot}) = 1$ for any overall multiplicity
distribution. Analogously, the individual terms in the
cumulant functions are divided by their respective integrals, so
that, for example,
\beql{ele}
C_3(\bm{x}_1,\bm{x}_2,\bm{x}_3)
\, \longrightarrow \,
c_3(\bm{x}_1,\bm{x}_2,\bm{x}_3) =
h_3 - \sum_{(3)} h_1 h_2 + 2 h_1 h_1 h_1
\eeq
yielding after normalization ({\it cf.\/} Eq.\ \req{cupb})
\beql{elf}
K_3(\epsilon) =
\eeq
\beqs
\lg {1\ov \hat N_a} \sum_{i_1}
\left(
        {  a^{[2]} \ov (\hat N_a - 1)^{[2]}  }
          - \lg b^{[2]} \ov \hat N_b^{[2]}  \rg
          - 2 { a \ov \hat N_a - 1} \lg b \ov \hat N_b \rg
          + 2 \lg b \ov \hat N_b \rg^2
\right)
\rg
\left/
\lg {1\ov \hat N_a} \sum_i \lg b \ov \hat N_b \rg^2 \rg
\right. \!\! .
\eeqs
Statistical independence is understood for these cumulants to mean a
factorization of the {\it probabilities} $h_q$ rather than of
the densities $\rho_q$.

\section{Bose-Einstein moments and cumulants}

One great advantage of correlation integrals in general is that
they allow the use of variables which are functions of two or more
particles \ct{Egg93d}, while conventional
binning is usually done in terms of single-particle variables only.

Bose-Einstein correlations are a prime example of the use
of relative coordinates: the quantum mechanical interference of
identical particles manifests itself in a rise of the
two-particle correlation function
\beql{qud}
k_2(\bm{p}_1,\bm{p}_2) =
{ \rho_2(\bm{p}_1,\bm{p}_2) \ov \rho_1(\bm{p}_1) \rho_1(\bm{p}_2) }
- 1
\eeq
at small relative momenta $\bm{q} = \bm{p}_1 {-} \bm{p}_2$.
Other formulations \ct{Gol60a}
test the correlation in terms of the one-dimensional
variable $Q^2 = -(p_1 - p_2)^2$, the relative four-momentum.
The correlation integral formalism can be utilized for both these
variables to yield moments and cumulants of higher order \ct{Egg93d}.
Taking $Q^2$ as an example, one first integrates out the
unneeded degrees of freedom in both $\rho_2$ and the normalization
$\rho_1\rho_1$ \ct{Ber77a},
\beqa
\lleq{qui}
\rho_2(Q^2) &=& \int d^3\bm{p}_1\, d^3 \bm{p}_2\,
\rho_2(\bm{p}_1,\bm{p}_2)\, \delta[Q^2 + (p_1-p_2)^2] \,,
\\
\lleq{qul}
\rho_1\!\otimes\!\rho_1(Q^2) &=& \int d^3 \bm{p}_1 \, d^3 \bm{p}_2 \,
\rho_1(\bm{p}_1)  \rho_1(\bm{p}_2) \,
\delta[Q^2 + (p_1-p_2)^2] \,,
\eeqa
which, using the delta function form of Eq.\ \req{ciaa},
translates into the measurement prescriptions
\beqa
\lleq{quj}
\rho_2(Q^2) &=& \lg \sum_{i\ne j}
\delta[Q^2 - (Q_{ij}^{aa})^2] \rg \,,
\\
\rho_1\!\otimes\!\rho_1(Q^2) = {1\ov\Nev^{[2]}} \sum_{a \neq b} \sum_{i,j}
\delta[Q^2 - (Q_{ij}^{ab})^2]
&=&
\lg \sum_i \lg \sum_j \delta[Q^2 - (Q_{ij}^{ab})^2] \rg \rg
\,,
\eeqa
where $(Q_{ij}^{ab})^2 = -(P_i^a - P_j^b)^2$ measures the
relative four-momentum between particles $i$ and $j$ taken from
two different events $a$ and $b$. Here, too, we see the
direct emergence of the event mixing prescription as the
appropriate method of normalizing Bose-Einstein correlation
functions.

For measurement of higher orders, one must first make an ansatz
how the $q$ three-momenta are to be combined into a single variable,
the choice of which depends on physical arguments of the
specific system and the signal being sought. One possibility is
to sum all $q(q{-}1)/2$ pairs of relative four-momenta
to give a measure of the overall $q$-particle four-momentum
\ct{Jur89a}, e.g.\ for $q=3$,
\beql{quii}
Q^2 = -(p_1-p_2)^2 - (p_1-p_3)^2 - (p_2-p_3)^2 \,;
\eeq
this amounts to a GHP-type topology of the correlation integral
in the four-momenta \ct{Egg93a,Lip92a}.
(The $Q^2$ defined in this way is
merely the $q$-particle invariant mass minus a constant.)
Moments are found by formulas analogous to Eq.\ \req{qui} above,
while cumulants are constructed directly from Eqs.\ \req{cua}{\it ff.}
inserted into
\beql{qve}
C_q(Q^2) = \int d^3\bm{p}_1\ldots d^3\bm{p}_q\,
C_q(\bm{p}_1,\ldots,\bm{p}_q)
\delta[Q^2 + \sum_{\alpha < \beta = 1}^q (p_\alpha - p_\beta)^2]
\,,
\eeq
i.e.\ the expansion of $C_q$ in terms of the $\rho_q$ must
be done before projection of the three-momenta onto $Q^2$.

\section{Biased and unbiased estimators}
\label{sec:ubia}

The use of the Star integral (or other forms such as the
form used above for Bose-Einstein correlations) permits much
more accurate measurements and hence will likely reveal more detailed
structure of the underlying dynamics. Greater accuracy requires,
however, that possible biases be understood on a higher level than
before. One such bias arising generally in the measurement of
correlations has to do with the theory of estimators \ct{Egg93ip}.

To understand this, we must go back to the basics of sampling theory.
For a given quantity of interest (``statistic'') $U$, there
ideally exists an infinite set of measurements $\hat U$;
this is termed the {\it population} of such measurements.
A statistical average based on the whole population would yield the
``true'' value $\bar U$ of this quantity.

In practice, the size of the set of measurements carried out is limited,
corresponding to a single {\it sample} of $\Nev$ measurements taken out of
the total population.
Many such samples $\cal N$ could theoretically be taken, each one yielding a
sample average $\langle U \rangle$, the set of which in itself forms a
distribution, the {\it sampling distribution}.
While there is no way to ascertain where
within this distribution the $\langle U \rangle_s$ obtained from a particular
sample will fall, at least one can test whether the average of this
sampling distribution coincides with $\bar U$. Surprisingly, such
a {\it sampling average}
\beql{exx}
\sav{U} = \lim_{{\cal N} \to \infty}
\sum_{s} \langle U \rangle_s / {\cal N}
\eeq
does not necessarily coincide with $\bar U$ except in the (for the
experimentalist uninteresting) case $\Nev \to \infty$.
When it does not, one looks for a modification, say $e(U)$;
which is called an {\it unbiased estimator} of $\bar U$ if it
fulfils the condition
\beql{uea}
\sav{e(U)} = \bar U \qquad \mbox{for all values of $\Nev$}\,.
\eeq
The age-old problem of finding suitable estimators has been extensively
investigated and we merely quote the results. It has been shown that
the inclusive density $\rho_q$ we have been using in the previous
sections is an unbiased estimator,
\beql{ueb}
 \sav{ \rho_q(\bm{x}_1,\ldots,\bm{x}_q) }  =
   \bar\rho_q(\bm{x}_1,\ldots,\bm{x}_q) \,;
\eeq
in addition, we note that the sampling average of a {\it single event}
inclusive density $\hat\rho_q$ as defined in Eq.\ \req{ciaa} also
yields $\bar\rho$ since Eq.\ \req{uea} is valid for samples
consisting of single events, $\Nev = 1$,
\beql{uec}
\sav{  \hat\rho_q(\bm{x}_1,\ldots,\bm{x}_q)  } =
       \bar\rho_q(\bm{x}_1,\ldots,\bm{x}_q) \,.
\eeq
However, whenever two or more event averages are involved,
the naive product of sample densities yields a biased estimator,
for example
$ \sav{ \rho_1(\bm{x}_1)    \rho_1(\bm{x}_2) } \neq
    \bar\rho_1(\bm{x}_1)\bar\rho_1(\bm{x}_2)
$
and
$ \sav{ \rho_2(\bm{x}_1,\bm{x}_2)    \rho_1(\bm{x}_3) }  \neq
    \bar\rho_2(\bm{x}_1,\bm{x}_2)\bar\rho_1(\bm{x}_3)
$,
so that all normalizations and higher-order cumulants discussed in
the previous sections must be corrected to yield unbiased estimators
of their corresponding expectation values.

Consider for example the product of two single-particle densities.
Let $A$ be the number of events used for averaging; when one corrects
interferometry or conventional factorial moments, $A$ will be equal to $\Nev$;
for the Star integral, the usual choice is $A = \Nev-1$ or, when a faster
inner event loop is desired, $A$ can be made much less than $\Nev$
\ct{Egg93a}. Now, using Eq.\ \req{aau}, we have
\beqs
\sav{ \rho_1(\bm{x}_1)    \rho_1(\bm{x}_2)}  =
\sav{ {1\ov A^2} \sum_{e_1 e_2}
         \hat\rho_1^{e_1}(\bm{x}_1)\hat\rho_1^{e_2}(\bm{x}_2)
    }
\neq
\sav{ {1\ov A} \sum_{e_1} \hat\rho_1^{e_1}(\bm{x}_1) }
\sav{ {1\ov A} \sum_{e_2} \hat\rho_1^{e_2}(\bm{x}_2) } ,
\eeqs
the second part being the desired true value
$\bar\rho_1(\bm{x}_1)\bar\rho_1(\bm{x}_2)$.
The reason why $\sav{\rho_1\rho_1}$ does not yield
the true value lies in the ``diagonal terms'' $e_1 = e_2$ in the
double sum above which prevent the desired factorization,
as $\sav{\hat\rho_1^{e_1}\hat\rho_1^{e_2}} \neq
\sav{\hat\rho_1^{e_1}} \sav{\hat\rho_1^{e_2}}$ unless $e_1$ and $e_2$
refer to two different (and hence independent) events.
Clearly, the desired unbiased estimator is given by
the double sum restricted to unequal events, since
\beql{ueg}
\sav{ {1\ov A(A{-}1)} \sum_{e_1 \neq e_2}
         \hat\rho_1^{e_1} \hat\rho_1^{e_2}
    }
=
{1\ov A(A{-}1)} \sum_{e_1 \neq e_2} \sav{ \hat\rho_1^{e_1} \hat\rho_1^{e_2} }
=  \sav{\hat\rho_1^{e_1}} \sav{\hat\rho_1^{e_2}}
= \bar\rho_1 \bar\rho_1 \,.
\eeq
In general, therefore, the unbiased estimator for the product of $q$
single-particle densities is given by
\beql{ueh}
{1\ov A^{[q]}} \sum_{e_1 \neq e_2 \neq \ldots \neq e_q}
\hat\rho_1^{e_1}\hat\rho_1^{e_2} \cdots \hat\rho_1^{e_q} \,.
\eeq
{}From this prescription, we obtain the unbiased estimators
for the Star integral normalization \ct{Egg93ip},
\beqa
\hat\xi_2^{\rm norm} &=& \lgs b \rgs \,, \\
\hat\xi_3^{\rm norm}
&=& \lgs b \rgs^2 - {\kappa_2(b,b) \ov (A-1)}   \,, \\
\hat\xi_4^{\rm norm}
&=& \lgs b \rgs^3 - {3 \lgs b \rgs \kappa_2(b,b) \ov (A-1)}
     + {2\kappa_3(b,b,b) \ov (A-1)^{[2]} } \,,
\\
\hat\xi_5^{\rm norm}
&=& \lgs b \rgs^4
   - {6 \lgs b \rgs^2 \kappa_2(b,b) \ov (A-1)   }
   + {  8 \lgs b \rgs \kappa_3(b,b,b)  +  3 \kappa_2^2(b,b)
       \ov  (A-1)^{[2]}  } \non\\
&&\mbox{} \quad\quad
   - { 6\kappa_4(b,b,b,b) + 9 \kappa_2^2(b,b)  \ov (A-1)^{[3]}  }
\,,
\eeqa
where
\beqa
\lleq{ubv}
\kappa_2(U,V) &=& \lgs UV  \rgs - \lgs U \rgs \lgs V \rgs  \,, \\
\kappa_3(U,V,W) &=& \lgs UVW \rgs
         - \sum_{(3)} \lgs UV \rgs \lgs W \rgs
         + 2 \lgs U \rgs \lgs V \rgs \lgs W \rgs         \,,  \\
\kappa_4(U,V,W,X) &=& \lgs UVWX \rgs
           - \sum_{(4)} \lgs U \rgs \lgs VWX \rgs
           - \sum_{(3)} \lgs UV \rgs  \lgs WX \rgs     \non \\
&&
\mbox{}  + 2\sum_{(6)} \lgs U \rgs \lgs V \rgs \lgs WX \rgs
           - 6 \lgs U \rgs \lgs V \rgs \lgs W \rgs \lgs X \rgs  \,,
\eeqa
where the sums indicate the number of combinations to be taken
and $U, V,\ldots $ is any statistic of interest; for example
$\kappa_2(b,b^{[2]}) = \langle b b^{[2]} \rangle
- \langle b \rangle \langle b^{[2]} \rangle$.
Note that (for the Star integral)
the second order normalization does not need a
correction; this is because the first event sum over $e_1$ is
always pulled out in calculating the $a$-quantities; only the
sums over $e_2, e_3 \ldots$ must be made explicitly unequal.

For the cumulants, we need the more general statement:
if $\rho_{q_1}$, $\rho_{q_2}\ldots \rho_{q_K}$
are densities of order $q_1, q_2, \ldots q_K$, the unbiased estimator
of their product is given by
\beql{uei}
{1\ov A^{[K]}} \sum_{e_1 \neq e_2 \neq \ldots \neq e_K}
\hat\rho_{q_1}^{e_1}\hat\rho_{q_2}^{e_2} \cdots \hat\rho_{q_K}^{e_K} \,.
\eeq
Implementing these, we find the unbiased estimators for the
$i$-particle cumulants to be
\beqa
\hat\rmf_2(i) &=& a -    \lgs b       \rgs     \,, \\
\hat\rmf_3(i)
&=& a^{[2]} -            \lgs b^{[2]} \rgs
            - 2 a        \lgs b       \rgs
            + 2          \lgs b       \rgs^2
            - {2\ov A-1} \kappa_2(b,b)
\,, \\
\hat\rmf_4(i)
&=& a^{[3]} -            \lgs b^{[3]} \rgs
            - 3 a^{[2]}  \lgs b       \rgs
            - 3 a        \lgs b^{[2]} \rgs
            + 6          \lgs b       \rgs     \lgs b^{[2]} \rgs
            + 6 a        \lgs b       \rgs^2
            - 6          \lgs b       \rgs^3
\non\\
& &\mbox{}
+ {6\ov A-1} \left\{  \left( 3 \lgs b \rgs - a \right) \kappa_2(b,b)
                    - \kappa_2(b,b^{[2]})
             \right\}
- {12\ov (A-1)^{[2]} }\kappa_3(b,b,b)
\,, \\
\hat\rmf_5(i)
&=& a^{[4]} -            \lgs b^{[4]} \rgs
            - 4 a^{[3]}  \lgs b       \rgs
            - 4 a        \lgs b^{[3]} \rgs
   \non\\
& &\mbox{}  -  6 a^{[2]} \lgs b^{[2]} \rgs
            +  8         \lgs b       \rgs    \lgs b^{[3]} \rgs
            + 12 a^{[2]} \lgs b       \rgs^2
            +  6         \lgs b^{[2]} \rgs    \lgs b^{[2]} \rgs
\non\\
& &\mbox{}  + 24 a       \lgs b       \rgs    \lgs b^{[2]} \rgs
            - 36         \lgs b       \rgs^2  \lgs b^{[2]} \rgs
            - 24 a       \lgs b       \rgs^3
            + 24         \lgs b       \rgs^4
\non\\
& &\mbox{}
- {2\ov A-1} \left\{  \left(
                      6 a^{[2]}        - 18 \lgs b^{[2]} \rgs
                   - 36 a \lgs b \rgs  + 72 \lgs b       \rgs^2
                      \right) \kappa_2(b,b)
             \right.
\non\\
& &\mbox{}   \left. \qquad\qquad\qquad
                   + 4 \kappa_2(b      ,b^{[3]})
                   + 3 \kappa_2(b^{[2]},b^{[2]})
                   + \left( 12a - 36 \lgs b \rgs
                        \right) \kappa_2(b,b^{[2]})
             \right\}
\non\\
& &\mbox{}
+ {24\ov (A-1)^{[2]}}
           \left\{ 3 \kappa_2^2(b,b)
                 + (8 \lgs b \rgs - 2a) \kappa_3(b,b,b)
                 - 3 \kappa_3(b,b,b^{[2]})
           \right\}
\non\\
& &\mbox{}
- {72\ov (A-1)^{[3]}}
           \left\{ 2 \kappa_4(b,b,b,b) + 3 \kappa_2^2(b,b)
           \right\}
\,.
\eeqa
For very small samples, when the inner event average sum $\sum_b$ cannot be
taken strictly over $b\neq a$, corrections must also be made for
equal-event terms \ct{Egg93ip}. These are very important for small
samples found e.g.\ in fixed-$N$ cuts and cosmic ray data.
When full event mixing is implemented for Bose-Einstein correlations
and conventional factorial moments or cumulants, similar bias
corrections are mandatory \ct{Egg93ip}.

\vspace{2mm} \noindent
{\bf Acknowledgements:} HCE gratefully acknowledges support by the
Alexander von Humboldt Foundation.


\end{document}